# Time-spliced X-ray Diffraction Imaging of Magnetism Dynamics in a NdNiO$_3$ Thin Film

Kenneth R. Beyerlein[*]

Max Plank Institute for the Structure and Dynamics of Matter, Hamburg, 22767, Germany

*Corresponding Author: kenneth.beyerlein@mpsd.mpg.de; t: +49 40 8998 5787,
ORCID: 0000-0002-7701-4369

ABSTRACT

Diffraction imaging of non-equilibrium dynamics at atomic resolution is becoming possible with X-ray free-electron lasers. However, there are unresolved problems with applying this method to objects that are confined in only one dimension. Here I show that reliable one-dimensional coherent diffraction imaging is possible by splicing together images recovered from different time delays in an optical pump x-ray probe experiment. The time and space evolution of antiferromagnetic order in a vibrationally excited complex oxide heterostructure is recovered from time resolved measurements of a resonant soft X-ray diffraction peak. Mid-infrared excitation of the substrate is shown to lead to a demagnetization front that propagates at a velocity exceeding the speed of sound, a critical observation for the understanding of driven phase transitions in complex condensed matter.

SIGNIFICANCE STATEMENT

Imaging the atomic scale dynamics of thin films is important to develop the next generation of computer technology. Coherent diffraction imaging can provide this information for other dimensionalities, but is unreliable when applied to thin film measurements. This paper describes a new approach to solving this problem using many measurements on a system that is changing in time. As an example, a demagnetization front is imaged as it sweeps through an antiferromagnetic film at twice the speed of sound, leaving a paramagnetic state in its wake. This fast switching is initiated by a mid-infrared pulse tuned to the substrate. The recovered magnetization evolution then shows the potential for control of opto-electronic switching devices by driving interface lattice dynamics.

/body

I. INTRODUCTION

A growing trend in science and technology involves the use of advanced imaging techniques to study the non-equilibrium evolution of matter (1–3). X-ray free-electron lasers promise a full mapping of ultrafast dynamics, which can be achieved by time-resolved coherent X-ray diffraction imaging (4–6). One such approach employs iterative projection algorithms to recover the lost phase information, and thereby a real space image, from the oversampled diffraction intensity of a compact object (7–9). However, this phase retrieval problem is only generally unique for two or three dimensions.

In one dimension, phase retrieval for a convex object is unique if all of the zeros of its Fourier modulus are real (10, 11). In fact, one-dimensional phase retrieval has been reported to reliably recover the strain profile of a thin film (12), density fluctuations near thin surfaces or interfaces (13–15), the amplitude and phase of ultrashort optical pulses (16), and the spatial distribution of the critical current in a superconducting Josephson junction (17–19). Uniqueness can be enforced if the phase of a few data points in the time domain are known (20), or if two different signals are interfered before being recorded (21). In most physical cases, the number of complex zeros is at least finite, leading to a countable number of possible solutions (10, 22, 23). It is not possible to determine if a unique solution exists by looking at the measured intensity distribution alone. Multiple trials of a phasing algorithm with different random initial conditions are necessary to test the fidelity of the recovered image (7, 8). If multiple solutions exist, the algorithm will recover each of the possible solutions, thus the problem becomes determining which of the recovered images could be the correct representation of the object.

The present manuscript proposes determining the correct image by attempting to splice together a movie of an object that is evolving in response to an experimental parameter. The solution set is then constrained by finding the series of images that show a logical causal progression. This approach is demonstrated by recovering the light-induced magnetic order dynamics in a correlated electron thin film from time-resolved resonant soft x-ray diffraction measurements. In this case, it affirms the recovered solutions and resolves ambiguities in the relative object placement between frames. In some ways, the added time constraint is analogous to that imposed when recovering the wave form of ultrashort optical pulses from frequency-resolved optical gating (FROG) (24–27). However, the difference is that the FROG spectrogram represents multiple time-windowed measurements of a single light pulse, while time-splicing uses similarities found in independent observations of an evolving object.

Phase transitions in materials, especially those with strong electronic correlations, often involve intertwined atomic, electronic, and spin degrees of freedom. Systematic time-resolved resonant X-ray diffraction studies have probed the evolution of the *spatial average* of these factors in response to optical stimuli (28–31). However, little is understood about the non-equilibrium spatial evolution of these driven transitions. One interesting example of such a spatially heterogeneous transformation is the reported demagnetization and electronic perturbative fronts launched at the interface of an $LaAlO_3$/$NdNiO_3$ (LAO/NNO) heterostructure after mid-infrared excitation (32–34). The demagnetization front induced by lattice excitation of the substrate was found to travel at supersonic speed, suggesting electronic origins. However, this result came from fitting a time series of Bragg peak profiles to a model for the front propagation that could have introduced bias. To test this result, I have applied coherent diffraction imaging to recover the magnetic order dynamics in the film without any *a priori* assumptions.

The noncollinear antiferromagnetic (AFM) structure of $NdNiO_3$ in the insulating state has been determined by resonant soft X-ray diffraction (RSXD) (35, 36) and is shown in Figure 1. In this structure, the magnetic moments at the Ni sites within (111) planes are aligned along either the [111] or [-1-12] directions. The alignment rotates by 90 degrees between consecutive planes, leading to a four-times-larger cubic superlattice. The quantum mechanical mechanism that stabilizes this unusual AFM structure is currently debated, but it is believed to be related to other dynamic exchange mechanisms, where spin hopping between Ni sites is mediated by the Ni-O-Ni bonding angle (37). The resulting (¼ ¼ ¼) diffraction peak has been shown to be solely of magnetic origin (36) without charge and orbital ordering contributions (38).

## II. MAGNETIC SCATTERING THEORY

For the experiment sketched in Figure 1, a 30-nm-thin $NdNiO_3$ film grown on a pseudo-cubic (111) $LaAlO_3$ substrate was cooled to 40K, well below its metal-to-insulator transition temperature of 130K. Resonant soft X-ray diffraction $\theta$-$2\theta$ scans of the (¼ ¼ ¼) AFM superlattice reflection were then made at the Ni L$_3$ edge using *p*-polarized 852 eV X-rays. A polarization analyzer was not placed in the diffracted beam, so a combination of *s*- and *p*-polarized X-rays were measured. Femtosecond time-resolved measurements of the magnetic order dynamics in the nickelate film were carried out at the SXR beamline of the Linac Coherent Light Source (32). Mid-infrared and near-infrared laser excitation was used to investigate the differences between substrate lattice-driven heterogeneous magnetization dynamics and homogeneous electronically driven dynamics. The 4-mJ/cm$^2$ mid-infrared (mid-IR) pump pulses were of 200 fs duration at 15 μm wavelength, which is resonant with an optical phonon of the LaAlO$_3$ substrate, but not NdNiO$_3$ (32). The 800-nm near-infrared pulses of equivalent fluence were 100 fs in duration. Further information about the sample and experiment has been detailed previously (32).

The measured intensity from magnetic scattering depends on the orientation of the X-ray polarization relative to the magnetic moments in the material. Starting from general equations describing the magnetic structure factor in a $\theta$-$2\theta$ geometry (39), expressions for the measured diffraction intensities from the incident *p*-polarized X-rays were derived, and further details are given in the SI Appendix. The in-plane alignment of the Ni moments allows the calculation to be reduced to the components along the scattering vector direction. The resulting structure factors for *p*-to-*s*-polarization ($F_{\pi\sigma}$) and *p*-to-*p*-polarization ($F_{\pi\pi}$) scattering from an ideal AFM NdNiO$_3$ unit cell are

$$F_{\pi\sigma} = 2mf_{Ni}(\sin\theta + i\cos\theta\cos\psi), \tag{1}$$

$$F_{\pi\pi} = 2mf_{Ni} i \sin 2\theta \sin\psi. \tag{2}$$

Here $m$ represents the average magnetic moment for Ni in a (111) plane, and $f_{Ni}$ is the Ni resonant atomic scattering factor, while $\theta$ and $\psi$ are the scattering angle and angle between the scattering plane and the [-1-12] direction, respectively. It is evident from these equations that changing the magnetic moment magnitude ($m$) will only influence the structure factor amplitude, while changing the orientation of the magnetic moments ($\theta$ and $\psi$) will influence their amplitude and phase.

If the average magnetization magnitude or orientation vary through the film, the structure factor also becomes a function of film depth, $f(z)$. The measured intensity profile, $I(q)$, from one of the reflected polarizations is then related to this structure factor profile by

$$I(q) = \left| \int_{-\infty}^{\infty} f(z) \, e^{2\pi i q z} \, dz \right|^2 \tag{3}$$

Then, iterative phase retrieval can be used to recover the amplitude and phase of the structure factor profile from the measured intensity. The present experiment is slightly more complicated as both *s*- and *p*-polarized reflected X-rays were measured. The intensity is then given by the incoherent sum of profiles from the two polarizations. As shown in the SI Appendix, assuming that only the average in-plane magnetization magnitude is dependent on depth, and not the magnetization orientation, a total structure factor profile can be defined that is related to the magnetization profile, $m(z)$, by

$$f(z) = 2N f_{Ni} \alpha(\theta, \psi) m(z) e^{i\phi(z)} \tag{4}$$

where

$$\alpha(\theta, \psi) = (\sin^2\theta + \cos^2\theta \cos^2\psi + \sin^2 2\theta \sin^2\psi)^{1/2} \tag{5}$$

and $N$ is the number of unit cells in a scattering volume. Equation (4) shows that the magnetization profile through the film is proportional to the amplitude of the recovered structure factor profile, as previously alluded to from inspection of the unit cell structure factors.

In our case, the physical significance of the recovered phase profile, $\phi(z)$, is clouded by the fact that both X-ray polarizations were measured (40), and that it can have many different contributing factors. Namely, the structure factor phase is shown in the SI Appendix to be related to the magnetization orientation, but it can also be influenced by strain gradients in the film (12, 33, 41). As separation of these contributions to the phase requires correlating multiple measurements at different polarizations and Bragg reflections, the following analysis focuses on the magnetization profile obtained from the amplitude of the recovered structure factor profile. Yet for completeness the recovered phase profiles for all presented data are shown in SI Appendix, Fig. S5 and S6.

III. TIME-SPLICING ALGORITHM

An algorithm has been developed to obtain these depth profiles by performing iterative phase retrieval on the rocking curve measurements and checking the solution set by splicing together the results from different time delays. The general steps of the analysis are sketched in Figure 2. First, the measured intensity was expressed in terms of the scattering vector magnitude, corrected for background, absorption and the Lorentz factor, and then converted to the diffraction modulus by taking its square root. The resulting modulus was used as input for a set of fifty independent phasing trials. This number was found to be a sufficient representation of the solution space as sets containing more trials were found to have the same standard deviation after alignment. A different starting point for each trial was made by combining the modulus with random initial phase values and taking its Fourier transform. The scattering factor and phase profiles were then refined using an error reduction (ER) algorithm (7, 42), that consisted of a modulus followed by a support projection operation. In all cases, 1000 iterations of this algorithm were enough to reach convergence of the residual metric defined by

$$R = \sum (A_{exp}(q) - A_{rec}(q))^2 / \sum A_{exp}(q)^2, \qquad (6)$$

where $A_{exp}(q)$ is the experimentally measured modulus and $A_{rec}(q)$ is the modulus of the reconstruction. Examples of the profiles obtained by averaging the converged trial results before alignment and splicing are shown in the central plot of Figure 2. The 5-nm resolution of the real space depth profile was given by the inverse of the measured range in reciprocal space and is indicted by the horizontal bars on the points shown in the figure. This is not the limit of the technique, as better resolution of the recovered profiles may be achieved by increasing the measured rocking curve angular range. A fixed support was found independently for each time delay by plotting the converged R-factor from phasing trials with different support sizes between 10 and 40 nm. The accepted support size was then taken as that corresponding to the bend in the resulting L-curve (43). Further details about the support size determination and examples of the obtained L-curves are given in the SI Appendix.

The different trial results for a given time delay were then aligned to correct for ambiguities in the solutions before averaging and assess the uniqueness of the solution set. The so-called trivial ambiguities for a complex profile, $f(z)$, are a translation and phase offset, $f(z + c)\exp(i\phi_c)$, combined with a mirror-conjugation, $f^*(-z + c)\exp(i\phi_c)$ (7). The mirror-conjugation operation corresponds to an ambiguity of the orientation of the magnetization profile relative to the interface, known as the problem of determining the direction of time in short optical pulse measurements (24).

Aligning the time delays was done by first correcting for the average phase offset of the profile contained within the support. Then, a seed solution was randomly chosen from the set for translation and orientation alignment. The best translational alignment for each solution was found by minimizing the real-space correlation function

$$r_i(a) = \sum_z |f_i(z-a) - f_s(z)|^2, \tag{7}$$

where $f_i(z)$ and $f_s(z)$ are the complex valued profiles of the $i$-th solution and the seed solution respectively. Then, a mirror-conjugated solution was generated, $f_i^*(-z)$, and aligned to the seed solution. This mirror solution replaced the original solution if it was found to have a lower minimum value of $r_i$. In this procedure, the use of the square modulus of the complex distance vector in Eqn. (7) ensures that the amplitude and phase of the recovered profiles were considered during alignment.

The uniqueness of the solution set was assessed by inspecting some of the individual trial results, as well as, calculating a standard deviation profile from the set. For each time delay, the trial solution set consisted of multiple copies of a smaller set of unique solutions. However, only slight variation was found between the unique solutions that seemed to be consistent with the noise level of the measurement. The rather uniform standard deviation found along the depth of the aligned profiles supports this, indicating well correlated solutions were found for all time delay scans that were solved. Therefore, these measurements did not appear to result in multiple conflicting solutions or stagnation, which might be a consequence of using a tight fixed support (44) or related to the rather simple monotonic variation of the magnetization along the profile. The accepted set of solutions for each time delay were then averaged.

Finally, these average solutions were spliced together, allowing to check for consistencies in the recovered solutions from different time delays. The average negative time delay solutions were taken as the initial seed profiles. Each positive time delay average profile was then taken in series and aligned within the preceding profile. This was done again using the complex distance error metric of Eqn (7). As shown in Figures 2 and 3, the region of the depth profile farther than 15 nm away from the interface was found to be rather similar between time delays. This supports the claim that a reliable solution was obtained for each time step. Furthermore, the standard deviation along the spliced profiles was reduced compared to the case of the average raw profiles because ambiguities in the position and orientation were overcome. While not needed for this case, if multiple distinct solutions had been recovered, each could be spliced into the time series, with the most likely solution being that which minimizes the real space error metric of Eqn. (7). However, further work is needed to understand the generality of this method. Namely, before it can be applied to multilayer heterostructures, it remains to be seen if it can overcome the non-uniqueness caused by a convex support. Furthermore, the implementation of more noise-robust iterative phase retrieval algorithms, such as the modified difference map algorithm (45), could improve its applicability to noisy data.

## IV. RESULTS AND DISCUSSION

Figure 3a shows the time evolution of the reconstructed AFM ordering in the NNO film for time delays up to 5.5 picoseconds after mid-IR excitation. Here the AFM order parameter depth profiles, $m_t(z)$, have been normalized to that found at negative time delay, $m_{t_0}(z)$, according to

$$\hat{m}_t(z) = m_t(z)/m_{t_0}(z). \tag{8}$$

It is seen that after only 1 ps, the AFM ordering at the interface has already disappeared. As time progresses, a front that destroys the AFM ordering propagates into the film. After 3 ps, this front stagnates at a distance of 15nm from the interface. The AFM ordering near the film surface (25 nm from the interface) is found to decrease by only 20% within the first 1.5 ps and then stagnate. As shown in SI Appendix, Fig. S2, the AFM ordering begins to slowly recover after 22 ps – first growing back in the paramagnetic region near the interface before uniformly increasing in the film. Figure S4 shows that the magnetization recovery rate near the interface is six times faster than near the free surface, suggesting that the excited magnetic states in these two regions of the film are fundamentally different. It is also worth noting that the transition region separating these two magnetization states, where there is a steep magnetization gradient in the film, is found to be consistently 10-nm thick during the front propagation, stagnation and recovery dynamics.

AFM order depth profiles after near-IR excitation are shown in Figure 3b. Within just 0.5 ps, the magnetic order is found to decrease uniformly over the entire film by 80%. This fast uniform reduction of the AFM ordering fits with a mechanism of uniform absorption that causes charge exchange and disruption of the electronic configuration. The recovered profiles for later time delays (SI Appendix, Fig. S3) recover uniformly through the film as shown in SI Appendix, Fig. S4. The recovery rate for the near-infrared excitation is comparable to that found near the free surface of the mid-IR excited film. Therefore, the magnetization dynamics near the surface of the film seem to be due to electronic excitation from absorption of the mid-IR radiation.

These dynamics are in remarkably good agreement with those obtained previously by fitting the diffraction data with an error function model of the AFM order in the film (32). However, some differences exist between the two results. For instance, in the present study, the reconstructed profiles show a decrease in the magnetization near the film surface, while this was not captured by the model chosen in the original analysis. As a result, some variation in the front position is found between the two results, but this is within the resolution of the measurements.

Figure 4 shows magnetic order isolines that map out the full evolution of the magnetic order after mid-infrared excitation. The velocity of the propagating front can be measured using the position in the film where the AFM order drops to zero. The corresponding front velocity is a reliable measure of the transformation propagation speed, as opposed to group velocity measurements that have led to previous claims of superluminal light (46). From the slope of this line in the first 3 ps, the AFM-to-PM phase front is found to propagate at twice the speed of sound in NNO (8,200 m/s), which is in good agreement with the velocity found from the previous modeling of the data (32). Since the original study on the magnetization dynamics, other measurements on this system have found that acoustic deformation of the lattice propagates from the interface through the film slower, and an electronic disturbance sweeps through faster than the demagnetization front (33). While the mechanism driving the magnetization dynamics is still unclear, it has recently been predicted that nonlinear optical phonon coupling can modulate the exchange interaction in oxides and quickly disrupt AFM order (47). Further X-ray measurements are planned and the characteristic lengths and gradients obtained from order-disorder profiles like those shown here may be used to discern the driving mechanism.

In conclusion, time-spliced diffraction imaging has allowed for the visualization of a supersonic demagnetization front propagating in an NdNiO$_3$ thin film. The constraints imposed by time-splicing allowed for validation of the recovered profiles, and are believed to be useful to overcome cases of non-uniqueness in one-dimensional phase retrieval. We note that splicing methods are also applicable beyond time resolved studies. As long as a transition is continuous as a function of a control parameter (e.g. magnetic field, temperature or pressure), the recovered profiles can be spliced together to add constraints to the image recovery.

## ACKNOWLEDGEMENTS

The author would like to thank Ian Robinson for initial discussions; Andrea Cavalleri and Michael Först for supplying the data, many valuable discussions and reviewing the manuscript; Jörg Harms for figure illustration assistance; as well as, Andrew Morgan and Kartik Ayyer for discussions about the algorithm and iterative phase retrieval. This work was funded by the European Research Council (ERC) grant 319286 (Q-MAC).

**Figures**

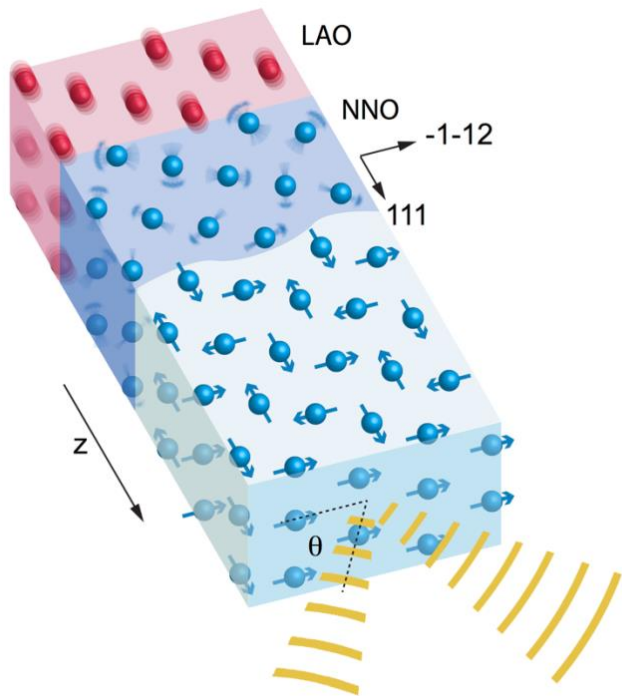

**Figure 1.** Illustration of the X-ray scattering geometry and magnetization dynamics after substrate resonant mid-IR excitation. Vibrations of the LAO lattice induced by mid-IR radiation lead to a demagnetization front propagating from the interface. The noncollinear AFM structure of Ni along the [111] direction of the NNO film is shown in the top surface of the unperturbed region. The diffraction measurements were made on the front surface of the film with the incident and exiting waves illustrated as yellow stripes and the $\theta$-scattering angle indicated. In this orientation of the NNO film, the $\psi$-angle is zero, as the projection of the incident X-ray beam into the surface plane is along [-1-12].

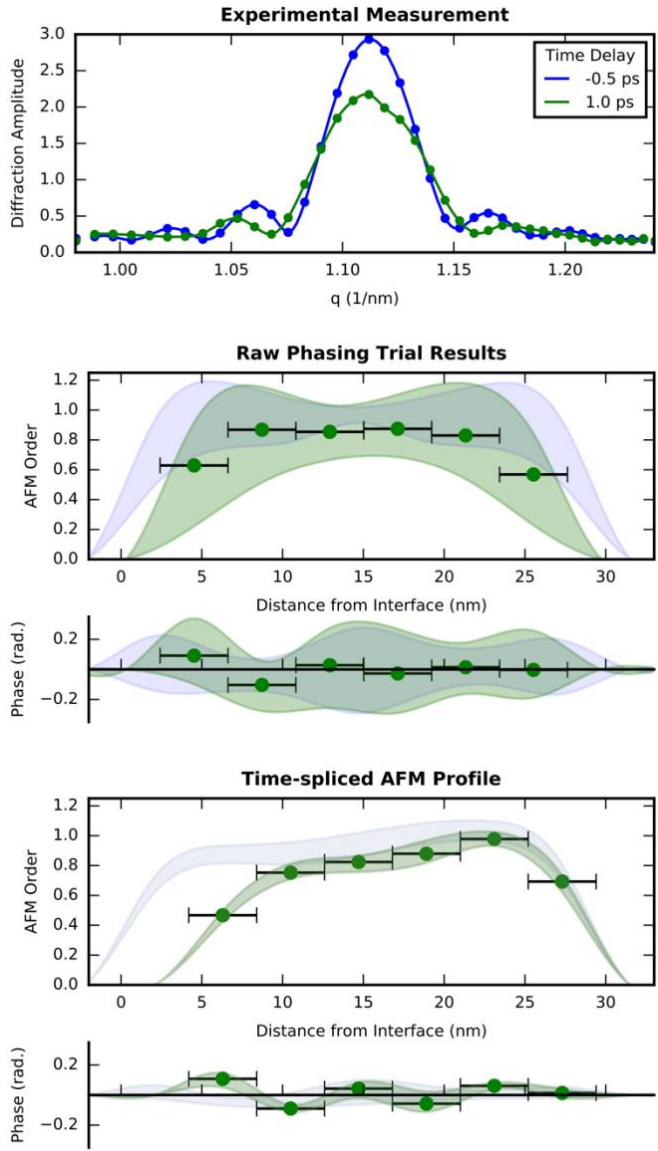

**Figure 2.** Workflow of the magnetization depth profile reconstruction algorithm. The first panel shows X-ray rocking curves measured with -0.5 and +1.0 ps mid-IR excitation time delays. For all time delays, 50 independent phasing trials were performed starting from the diffraction peaks like those shown. The average raw recovered AFM order and phase, profiles, $m(z)$ and $\phi(z)$, are shown in the central figure for the -0.5 ps (blue) and +1.0 ps (green) time delays. Here discrete values of the +1.0 ps average profiles are shown, with the horizontal error bars depicting the inverse q-range spatial resolution of the profile and the shaded areas depicting the standard deviation of the solution set. The raw set of solutions were then aligned and averaged before time-splicing. The lower figure then shows the final AFM order and phase profiles obtained after the average profiles are spliced together.

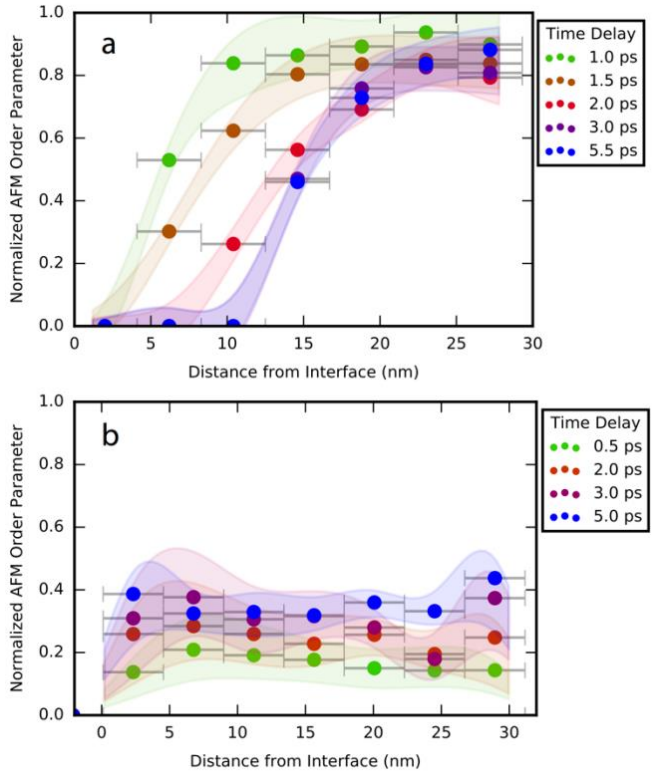

**Figure 3.** Temporal evolution of AFM ordering in the NNO film after optical excitation. (a) The AFM depth profiles normalized according to Equation (8) for time delays up to 6 ps after a mid-IR pump are shown. The points represent the average and the shaded area depicts the standard deviation from the aligned set of phasing trials. (b) The normalized AFM profiles of the NNO film after 800nm excitation are also shown.

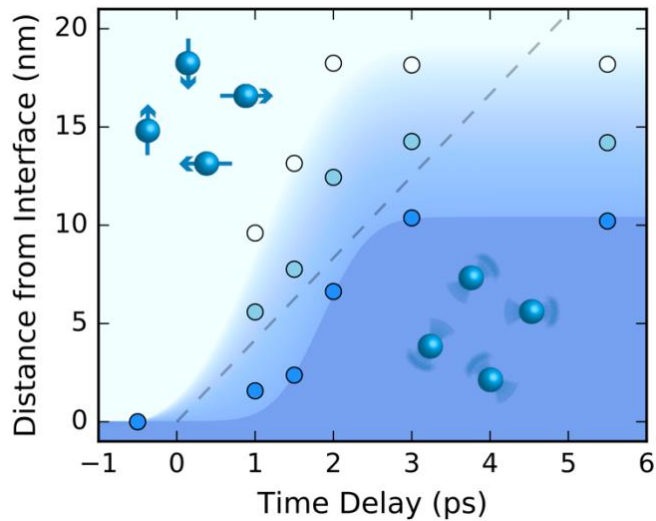

**Figure 4.** Evolution of the diffuse antiferromagnetic-to-paramagnetic front. The position of the initial anti-ferromagnetic ordering disturbance, inflection point of the front, and paramagnetic front were taken from the profiles in Figure 3a, and are shown as the white, light blue and dark

blue data points, respectively. Magnetization isolines that make up the background color gradient were then obtained by fitting error functions to such data. The color scale follows from that used for the data points. For reference, the propagation of a front travelling the speed of sound in NNO is depicted as a dashed line.

# SI Appendix

## I. Magnetic Scattering Theory

As measurements were made on the Ni L$_3$ absorption edge on the purely magnetic (¼ ¼ ¼) peak, we will only consider the contribution of electric dipole transitions to the magnetic cross section. From Hill and McMorrow (1), the atomic scattering amplitude is then given by

$$f_n = \left[(\hat{\varepsilon}' \cdot \hat{\varepsilon})F^{(0)} - i(\hat{\varepsilon}' \times \hat{\varepsilon}) \cdot \widehat{m_n}F^{(1)} + (\hat{\varepsilon}' \cdot \widehat{m_n})(\hat{\varepsilon} \cdot \widehat{m_n})F^{(2)}\right] \quad (S1)$$

where $\hat{\varepsilon}$ and $\hat{\varepsilon}'$ are the polarizations of the incident and scattered light, $\widehat{m_n}$ is the orientation vector of the magnetic moment of the n$^{th}$ ion, and $F^{(j)}$ are respective strengths of the three scattering processes. Again, as we are studying a magnetic reflection, the contribution of the first process will be zero. In a specular reflection geometry, this expression becomes Equation 15 of SI Appendix Reference 1. The atomic scattering factors for incident p-polarized light and scattered s- and p-polarized light are then respectively

$$f_{n,\pi\sigma} = \left[-iF^{(1)}(m_{n,3}\sin\theta - m_{n,3}\cos\theta) + F^{(2)}\left(m_{n,2}(m_{n,1}\sin\theta + m_{n,3}\cos\theta)\right)\right] \quad (S2)$$

$$f_{n,\pi\pi} = \left[-iF^{(1)}m_{n,2}\sin 2\theta - F^{(2)}\cos^2\theta\left(m_{n,1}^2\tan^2\theta + m_{n,3}^2\right)\right], \quad (S3)$$

and $\widehat{m_n} = m_{n,1}\hat{\imath} + m_{n,2}\hat{\jmath} + m_{n,3}\hat{k}$.

The Ni magnetic moments in the static AFM ordering of NdNiO$_3$ follows a ↑→↓← pattern along the [111] direction, where ↑ and ↓ are parallel to [111], while → and ← are along [-1-12]. In our scattering geometry, $\hat{k}$ is parallel to [111], and we will define $\hat{\imath}$ to be along [-1-12]. Then the magnetic moments for the four ions in the static magnetic unit cell are $\widehat{m_{\uparrow,\downarrow}} = \pm m\hat{k}$ and $\widehat{m_{\rightarrow,\leftarrow}} = \pm m\cos\psi\,\hat{\imath} \pm m\sin\psi\,\hat{\jmath}$, where $\psi$ is the angle between the incident wave vector projection into the plane and [-1-12].

The total scattering factor from a crystal for the π-to-σ scattering polarization is found from summing over all of the atomic layers according to

$$A_{\pi\sigma}(q) = \sum_n f_{n,\pi\sigma}\exp(-i2\pi q z_n), \quad (S4)$$

where $q$ is the magnitude of the scattering vector and $z_n$ is the position of the n$^{th}$ layer. The ionic position can be decomposed into the sum of the position a scattering cell, $z_p$, and the place of this layer within the cell, $z_r$, where $z_n = z_p + z_r$. The crystal scattering factor then takes the form

$$A_{\pi\sigma}(q) = \sum_p \sum_r f_{p,r,\pi\sigma}\exp(-i2\pi q z_r)\exp(-i2\pi q z_p) = \sum_p F_{p,\pi\sigma}\exp(-i2\pi q z_p). \quad (S5)$$

If the magnetic scattering cell is also allowed to deform by different amounts as a function of depth, the position of the p-th cell can be written as $z_p = pz_u + \Delta_p$, where $z_u$ is the ideal cell spacing and $\Delta_p$ is the displacement or strain of the p-th cell. This expression can be inserted into Eqn. (S5), resulting in

$$A_{\pi\sigma}(q) = \sum_p F_{p,\pi\sigma}\exp(-i2\pi q\Delta_p)\exp(-i2\pi q p z_u) \quad (S6)$$

This sum can be expressed as an integral over z, then the crystal scattering factor becomes

$$A_{\pi\sigma}(q) = \int F_{\pi\sigma}(z)\exp(-i2\pi q\Delta(z))\exp(-i2\pi q z)dz \quad (S7)$$

The same expression can be written for the ππ polarization amplitude, following the same arguments. The observed intensity is then the incoherent sum of the intensity from the two scattering polarizations,

$$I(q) = |A_{\pi\sigma}(q)|^2 + |A_{\pi\pi}(q)|^2. \tag{S8}$$

Summing over a single unit cell, using Equations (S2), (S3) and the magnetic moments defined previously, the expressions for the scattering factors of the ¼ ¼ ¼ reflection are

$$F_{\pi\sigma} = 2m f_{Ni}(\sin\theta + i\cos\theta\cos\psi), \tag{S9}$$
$$F_{\pi\pi} = 2m f_{Ni} i \sin 2\theta \sin\psi. \tag{S10}$$

In these expressions, we define the variable $f_{Ni}$ as equivalent to $F^{(1)}$ for nickel in the notation that was used in Equations (S2) and (S3) from SI Reference 1. The terms proportional to $F^{(2)}$ were found to cancel out for this reflection.

As the resolution in this experiment is larger than a unit cell, we will consider a larger scattering cell consisting of some number of unit cells, *N*. We will let the magnetization in each cell vary, leading to a z-dependence on the scattering factor as in Equation (S6). Now we will also describe the changes in magnetization in each scattering cell in terms of the projection of the average moment at each depth onto the static magnetization orientation. That is to say, we assume that the orientation angles of the moment ($\theta$ and $\psi$) are not z dependent. The scattering factor profiles then become

$$F_{\pi\sigma}(z) = 2N f_{Ni} m(z)(\sin\theta + i\cos\theta\cos\psi), \tag{S11}$$
$$F_{\pi\pi}(z) = 2N f_{Ni} m(z) i \sin 2\theta \sin\psi \tag{S12}$$

From Equation (S8), it is then found that the observed intensity becomes

$$I(q) = 4N^2 f_{Ni}^2 \alpha^2(\theta,\psi) |\int m(z)\exp(-i2\pi q z)dz|^2 \tag{S13}$$

and the scattering amplitude is

$$A(q) = 2N f_{Ni} \alpha(\theta,\psi) \int m(z)\exp(-i2\pi q z)dz \tag{S14}$$

where

$$\alpha(\theta,\psi) = (\sin^2\theta + \cos^2\theta\cos^2\psi + \sin^2 2\theta \sin^2\psi)^{1/2} \tag{S15}$$

If we ignore the incoherent sum in the iterative phase retrieval, we are treating the observed intensity originating from a single scattering factor profile as

$$I(q) = \left|\int_{-\infty}^{\infty} f(z) e^{2\pi i q z}\, dz\right|^2 \tag{S16}$$

and the amplitude is then

$$A(q) = \int_{-\infty}^{\infty} f(z) e^{2\pi i q z}\, dz. \tag{S17}$$

Then from equation S13, the obtained scattering factor profile is related to the magnetization profile by

$$f(z) = 2N f_{Ni} \alpha(\theta,\psi) m(z). \tag{S18}$$

In general, $f(z)$ is assumed to be a complex function, leading to the expression

$$f(z) = 2N f_{Ni} \alpha(\theta,\psi) m(z) e^{i\phi(z)}. \tag{S19}$$

So, $m(z)$ is proportional to the amplitude of the obtained scattering factor profile function recovered by iterative phasing algorithms. The phase profile $\phi(z)$ can be related to the strain profile, $\Delta(z)$, as shown in Equation (S7), however, next I will also show that under different assumptions it can also be related to the magnetization orientation profile.

Now assuming that the magnetization orientation angles are also dependent on the position in the film leads to a scattered intensity of the form

$$I(q) = 4N^2 f_{Ni}^2 \left[\left|\int m(z)(\sin^2\theta(z) + \cos^2\theta(z)\cos^2\psi(z))\exp(-i2\pi qz + i\phi_{\pi\sigma}(z))dz\right|^2 + \left|\int m(z)(\sin^2 2\theta(z)\sin^2\psi(z))\exp(-i2\pi qz + \pi/2)dz\right|^2\right] \quad (S20)$$

where $\phi_{\pi\sigma}(z) = \arctan(\cot\theta(z)\cos\psi(z))$.

While this is a more general treatment, the real space amplitude from each polarization is found to be a complicated product of three z-dependent functions. The separation of the recovered amplitude into these respective functions is not possible without multiple measurements at different incident beam $\psi$ angles for each time delay. Furthermore, applying phase retrieval to an incoherent sum of the amplitudes from the two polarizations will recover some complex average of them (2). Therefore, we will not consider this model as it does not allow for a clear physical interpretation of the results. As a corollary, we cannot differentiate from our analysis if the $m(z)$ profile recovered assuming Equation (S19) is truly due to a position dependence of the magnetic moment magnitude, or its orientation. Also, it becomes unclear if the recovered phase profile is due to a strain profile, or somehow related to $\phi_{\pi\sigma}(z)$ from Equation (S14).

## II. Iterative Phase Retrieval Projection Operators

An error-reduction algorithm was used to recover the amplitude and phase of the real space scattering profiles. One iteration of the algorithm updated the recovered the real space scattering profile, $f(z)$, by applying a reciprocal space modulus projection, $\boldsymbol{P_m}$, followed by a real space support projection, $\boldsymbol{P_s}$, as in

$$f_{n+1}(z) = \boldsymbol{P_s}\boldsymbol{P_m} f_n(z). \quad (S21)$$

The support projection is defined as

$$\boldsymbol{P_s} f_n(z) = \begin{cases} f_n(z) & \text{if } z \in S \\ 0 & \text{otherwise} \end{cases}, \quad (S22)$$

where $S$ is the set of points belonging to the support
The modulus projection was defined as

$$\boldsymbol{P_m} = \mathcal{F}^{-1}\widetilde{\boldsymbol{P}}_m \mathcal{F} f_n(z) \quad (S23)$$

with $\mathcal{F}$ denoting the fourier transform operation and $\widetilde{\boldsymbol{P}}_m$ defined as

$$\widetilde{\boldsymbol{P}}_m = A(q)e^{i\phi_n(q)} \quad (S24)$$

where $A(q)$ is the modulus of the measured intensity and $\phi_n(q)$ is the phase of $\mathcal{F}f_n(z)$. As described in the manuscript, this sequence of projections was repeatedly applied until convergence of $f(z)$ was reached.

## III. L-curve support determination

A fixed support size was determined for each time delay by conducting a trial of phase retrieval for different support sizes, $L$, and recording the converged values of the residual metric, $R$, defined in Equation (6). The support size was chosen as that corresponding to the bend in the

$R(L)$ curve. Examples of such curves for a few time delays are depicted in Figure S1. For the -0.5 ps time delay data, the support size of 30 nm was used, while for 5.5 ps a support size of 17 nm was used.

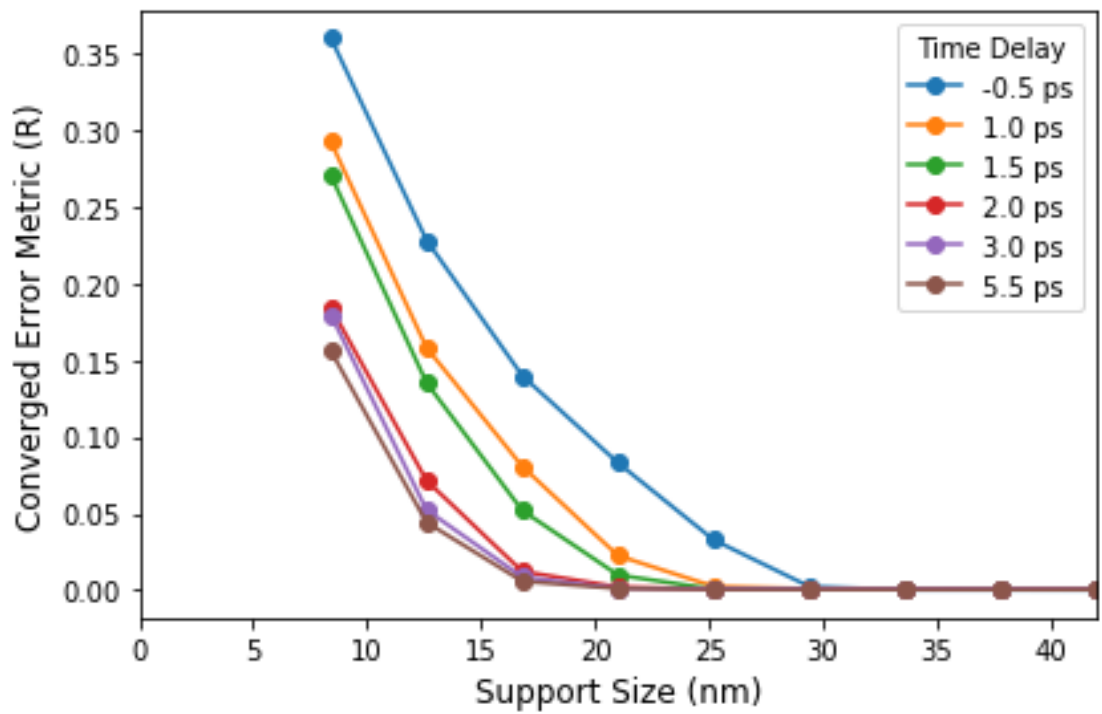

**Figure S1.** Trends of the converged values of the error metric assuming different support sizes.

## IV. Magnetization Recovery Dynamics

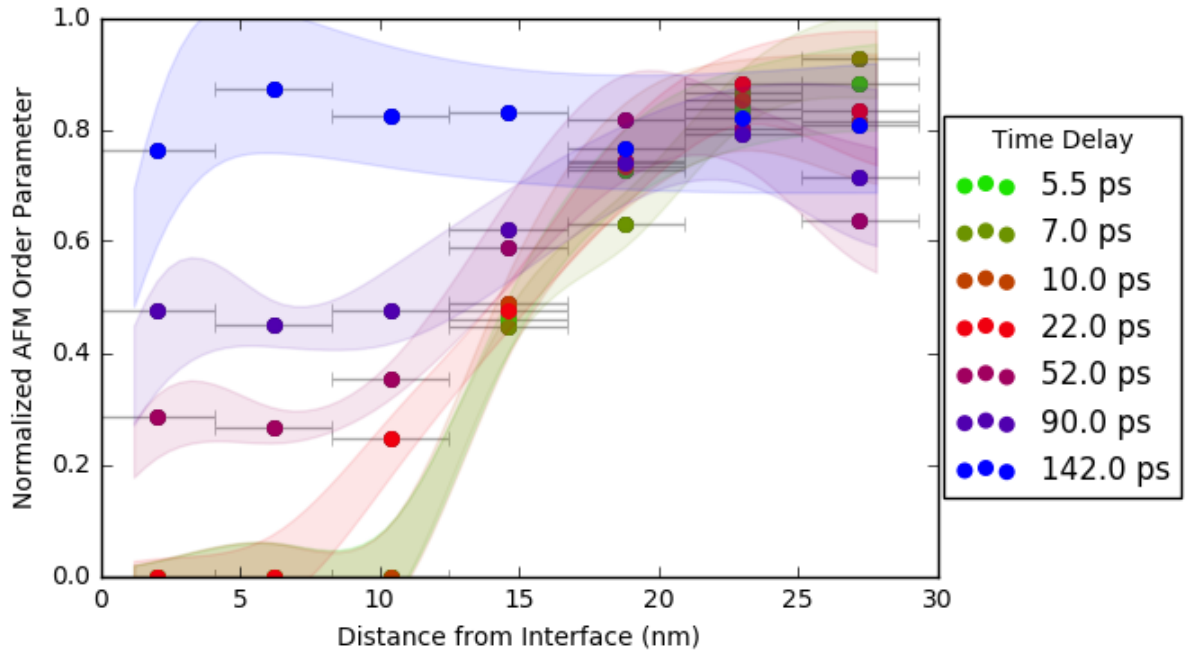

**Figure S2.** Magnetization profiles for long time delays after mid-IR excitation. Starting from the last time delay in Figure 3a, images of the AFM ordering recovery in the film after mid-IR excitation of the substrate are shown. The magnetization only begins to recover after 22 ps, and is nearly fully recovered after 142 ps.

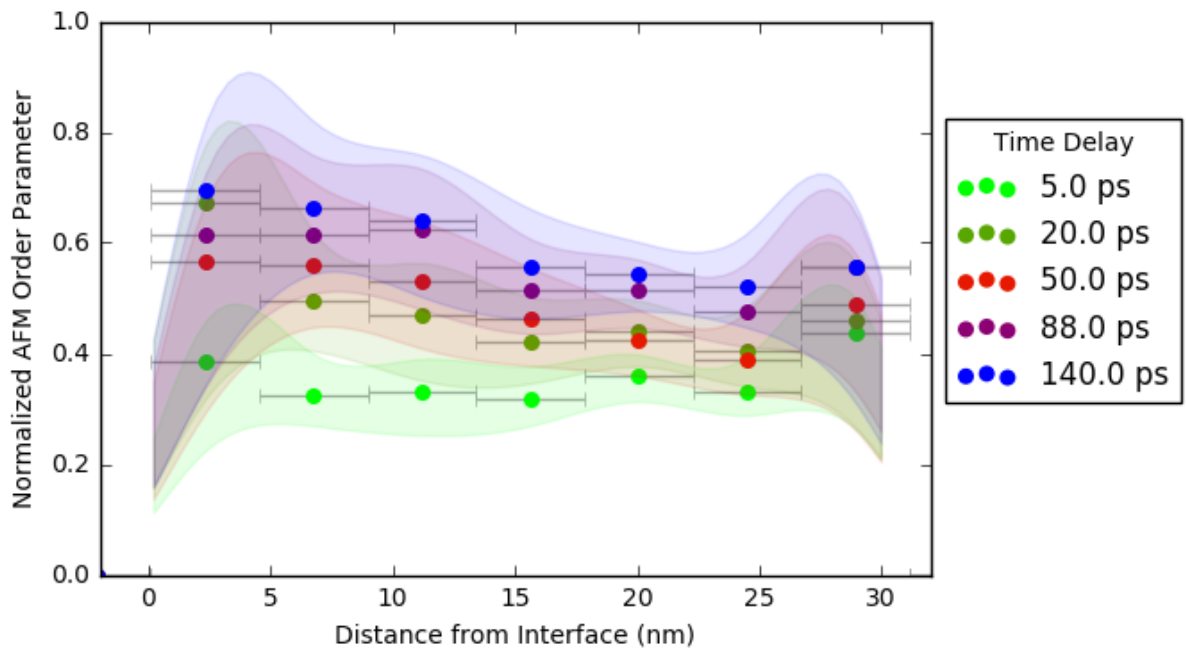

**Figure S3.** Magnetization profiles for long time delays after 800nm excitation. Starting from the last time delay in Figure 3b, images of the AFM ordering recovery in the film after 800nm

excitation of the film are shown. Even after 142 ps, the magnetization has only recovered to about 60% of its equilibrium value.

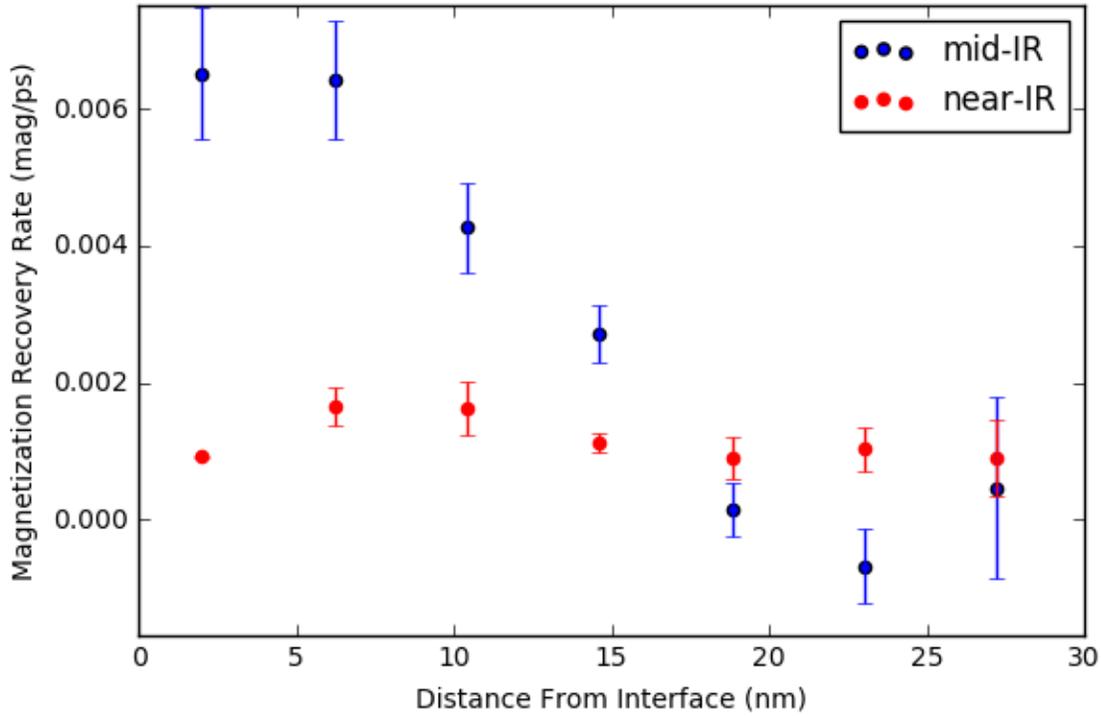

**Figure S4.** Rate of magnetization recovery as a function of position in the film for mid-infrared and near-infrared excitations. The normalized magnetization profiles from Figure S1 and S2 for time delays longer than 20 ps were analyzed to obtain the rate of recovery as a function of distance from the interface. Trends for this magnetization as a function of time at each position in the film were fit to a linear model, the slope of which is defined as the magnetization recovery rate.

## V. Recovered Phase Profiles

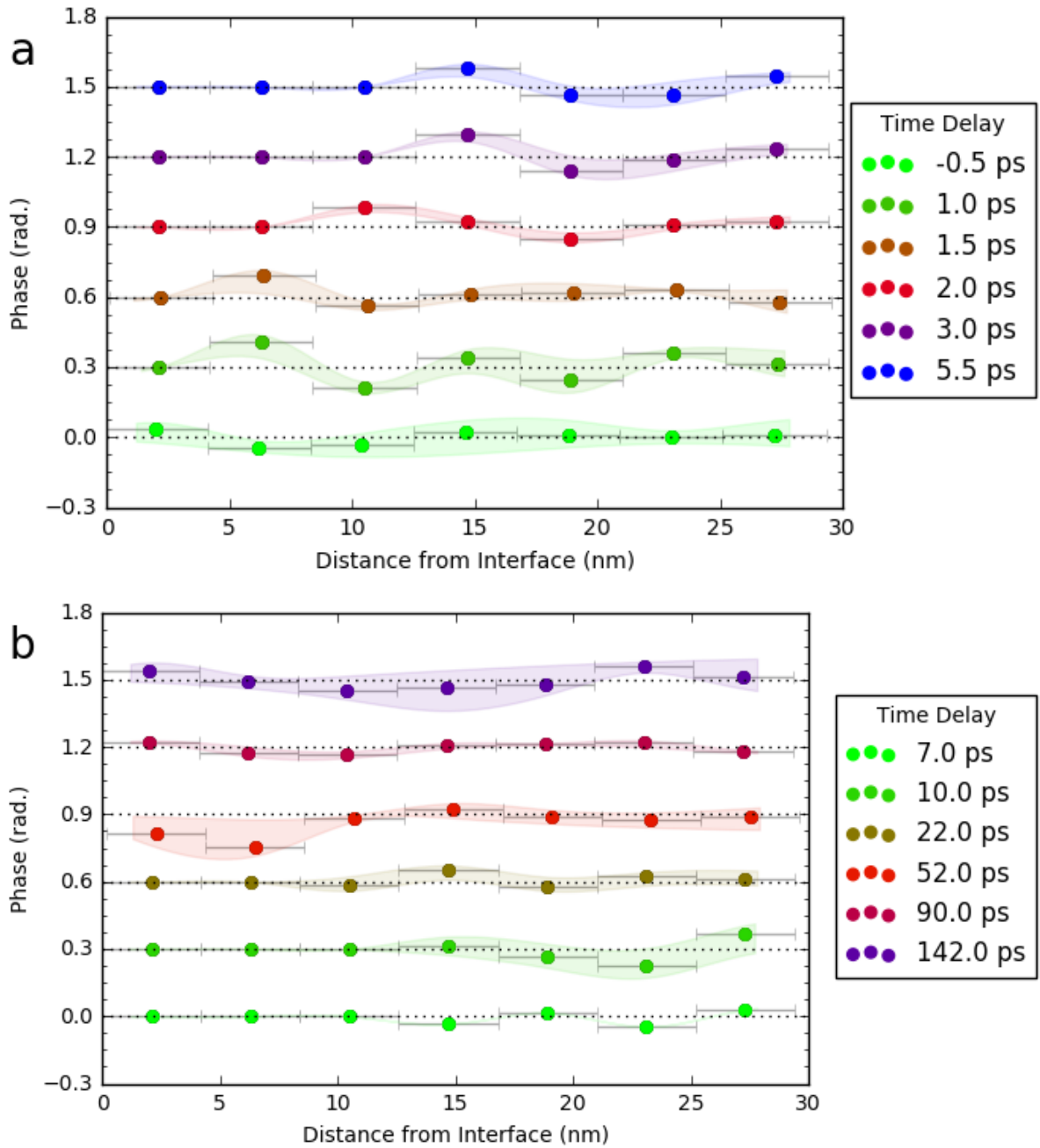

**Figure S5**. Recovered phase profiles for different time delays after mid-infrared excitation. Each profile has been offset for clear presentation and the dotted lines denote each respective zero phase line. (a) The phase profiles from Figure 3a, and (b) Figure S2 are during the early and later times of the magnetization recovery process are shown.

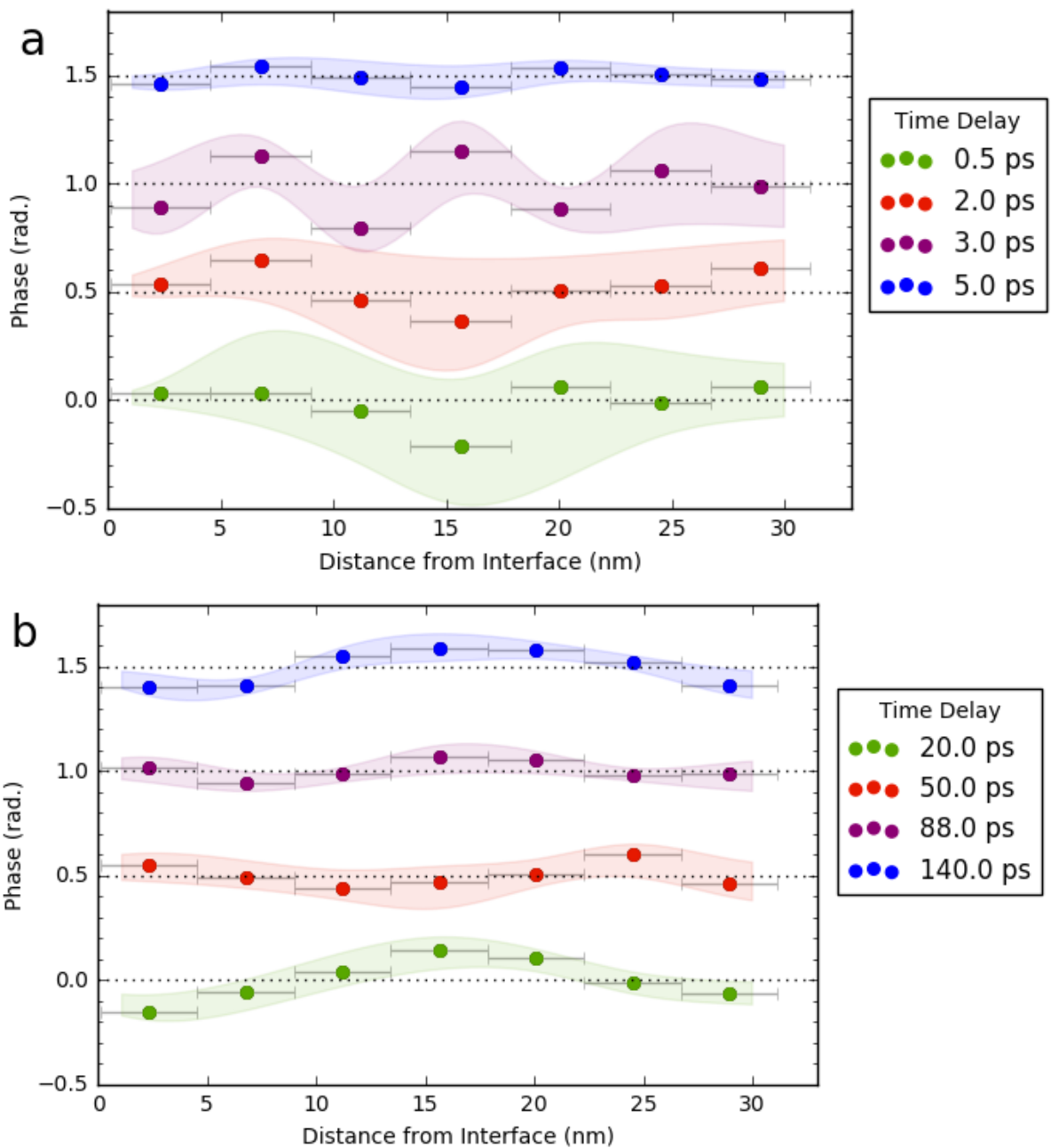

**Figure S6**. Recovered phase profiles for different time delays after near-infrared excitation. (a) The phase profiles from Figure 3b, and (b) Figure S3 are during the early and later times of the magnetization recovery process are shown.